\documentclass[conference]{IEEEtran}
\IEEEoverridecommandlockouts
\usepackage[utf8]{inputenc}
\usepackage{cite}
\usepackage{amsfonts,amssymb}
\usepackage{float}
\usepackage{amsmath}
\usepackage{mathtools}
\usepackage{lipsum}
\usepackage{amsmath,amssymb,amsfonts}
\usepackage{graphicx}
\usepackage{subcaption}
\usepackage{xcolor}
\usepackage{todonotes}
\usepackage{multirow}
\usepackage{soul}

\title{\LARGE \bf
Enhanced Rotational Invariant Convolutional Neural Network for Supernovae Detection
}

\author{\IEEEauthorblockN{Esteban Reyes\IEEEauthorrefmark{1}, 
Pablo A. Est\'evez\IEEEauthorrefmark{1}\IEEEauthorrefmark{2},
Ignacio Reyes\IEEEauthorrefmark{1}\IEEEauthorrefmark{2},
Guillermo Cabrera-Vives\IEEEauthorrefmark{3}\IEEEauthorrefmark{2},
Pablo Huijse\IEEEauthorrefmark{1}\IEEEauthorrefmark{2},\\
Rodrigo Carrasco\IEEEauthorrefmark{1}\IEEEauthorrefmark{2} and Francisco F\"orster\IEEEauthorrefmark{4}\IEEEauthorrefmark{2}}
\IEEEauthorblockA{\IEEEauthorrefmark{1}Department of Electrical Engineering Universidad de Chile, Chile\\
Email: esteban.reyes@ug.uchile.cl}
\IEEEauthorblockA{\IEEEauthorrefmark{2}Millennium Institute of Astrophysics, Chile} 
\IEEEauthorblockA{\IEEEauthorrefmark{3}Department of Computer Science, Universidad de Concepci\'on, Chile}
\IEEEauthorblockA{\IEEEauthorrefmark{4}Center for Mathematical Modeling 
Universidad de Chile, Chile}
}

\begin{document}

\maketitle

%%%%%%%%%%%%%%%%%%%%%%%%%%%%%%%%%%%%%%%%%%%

\begin{abstract}
In this paper, we propose an enhanced CNN model for detecting supernovae (SNe). This is done by applying a new method for obtaining rotational invariance that exploits cyclic symmetry. In addition, we use a visualization approach, the layer-wise relevance propagation (LRP) method, which allows finding the relevant pixels in each image that contribute to discriminate between SN candidates and artifacts. We introduce a measure to assess quantitatively the effect of the rotational invariant methods on the LRP relevance heatmaps. This allows comparing the proposed method, CAP, with the original Deep-HiTS model. The results show that the enhanced method presents an augmented capacity for achieving rotational invariance with respect to the original model. An ensemble of CAP models obtained the best results so far on the HiTS dataset, reaching an average accuracy of 99.53\%. The improvement over Deep-HiTS is significant both statistically and in practice.
\end{abstract}

%%%%%%%%%%%%%%%%%%%%%%%%%%%%%%%%%%%%%%%%%

\section{Introduction}

Astronomy is entering into a new era of big data due to the construction of very large scale facilities such as the Large Synoptic Survey Telescope (LSST), an 8.4 m telescope with a 3.2 Gigapixel camera, which will begin operations in northern Chile in 2022 \cite{huijse2014computational}. The LSST is a robotic telescope that will scan the entire southern hemisphere sky every 3 days, collecting information on 50 billion objects for 10 years \cite{ivezic2008lsst}. Time-domain astronomy studies stellar objects that change in time or position, e.g. supernovae (SNe), the explosive death of stars.  The High Cadence Transient Survey (HiTS) \cite{forster2016high} aimed at detecting SNe in their early stages in order to study the astrophysics associated with these phenomena. HiTS has a custom-made pipeline to process the images captured by the telescope and detect transients. Basically, the pipeline subtracts reference images from new images, detects sources and classifies them. The farther away from Earth, the higher the chance of finding SNe events because there are more galaxies. But deeper objects are usually fainter with a low signal-to-noise ratio. For this reason, among others, it is relevant to reduce significantly the false negative rate (FNR) and the false positive rate (FPR) at the output of this pipeline. In our previous work, we introduced a convolutional neural network (CNN) for classifying sources detected by the HiTS pipeline as true transients (`SN candidates') or bogus (`artifacts') \cite{cabrera2016supernovae}. In 2017 the model was enhanced by introducing partial rotational invariances, as well as improving the architecture and training algorithms, to yield the so-called Deep-HiTS model (DH) \cite{cabrera2017deep}. 

In this paper, we enhance Deep-HiTS by applying a new method for obtaining rotational invariance that exploits cyclic symmetry in CNNs \cite{dieleman2016exploiting}. In addition, we use a visualization approach, the layer-wise relevance propagation (LRP) \cite{samek2017evaluating}, in order to find the relevant pixels in each image that helps to discriminate between SN candidates and artifacts. We assess both qualitatively and quantitatively the effect of the rotational invariant methods using LRP and compare the original Deep-HiTS model with its enhanced version. In addition, we introduce ensemble classifiers to improve the performance of Deep-HiTS.

%%%%%%%%%%%%%%%%%%%%%%%%%%%%%%%%%%%%%%%%%

\section{Related Work}

\subsection{High Cadence Transient Survey and Deep-HiTS}

The High Cadence Transient Survey \cite{forster2016high} aims at detecting sky transients, ranging from hours to days, in particular it searches for early stage SNe. The data was collected from the Dark Energy Camera (DECam) \cite{flaugher2012status} mounted at the prime focus of the Victor M. Blanco 4 m telescope on Cerro Tololo near La Serena, Chile. The HiTS detection pipeline makes use of four images: \emph{template}, \emph{science}, \emph{difference} and \emph{signal-to-noise ratio} (SNR) \emph{difference} (\emph{difference} normalized by estimated image noise), see Fig. \ref{fig:CAP}. The \emph{difference} image is obtained from the subtraction between the \emph{template} (reference image) and \emph{science} (current image), in order to detect anything that changes in time or position. 

To generate the \emph{difference} images, the \emph{template} and \emph{science} must be aligned, which is achieved by applying the SExtractor object detector \cite{bertin1996sextractor} and then adjusting a second order transformation between objects detected in both images. To take into account changing atmospheric conditions that produce variations in brightness and blurring of images at the measurement time, a kernel point-spread-function (PSF) of the image with better conditions is matched to the worse one. After this PSF correction, the images are subtracted to obtain the \emph{difference}, which is normalized by its local noise to get the \emph{SNR difference} image. Transient candidates are those events with a local SNR greater than 5, and an image stamp of 21$\times$21 pixels is centered around each event. At the output of the HiTS pipeline, a classifier discriminates between true transients and bogus events. 

In a previous work, we proposed a CNN for discriminating between SN candidates and artifacts, the so-called Deep-HiTS model (DH) \cite{cabrera2017deep}. To be partially invariant to rotations, DH augmented the inputs by adding 90$^{\circ}$-180$^{\circ}$-270$^{\circ}$ rotations. Table \ref{table:architecture} shows the Deep-HiTS architecture. The inputs are 4 images of 21$\times$21 pixels. Then the rotation augmentation operation increases the batch size by 4. The first two convolutional layers have 32 filters of size 4$\times$4 and 3$\times$3, respectively. They are followed by a 2$\times$2 max-pooling layer. Next are 3 convolutional layers of 64 3$\times$3 filters, followed by a 2$\times$2 max-pooling layer. Up to this point,  each of the four rotations is processed separately. Just before the first fully connected (dense) layer, the feature maps for each rotation are flattened and concatenated in the feature dimension.  As a result, the first quarter of features corresponds to the feature maps with 0$^{\circ}$ rotations, the second quarter to 90$^{\circ}$ rotations, the third quarter to 180$^{\circ}$ rotations, and the last quarter to 270$^{\circ}$ rotations. The last operation transforms a minibatch of size 200 with feature maps of 6$\times$6$\times$64 to a batch of 50 original samples with a feature vector of size 9216 (=2304$\times$4) per sample. This feature vector passes through two dense layers of 64 neurons, ending with a dense softmax layer that generates the output probabilities for the one-hot encoding of the two classes. Leaky ReLUs are used as activation functions. A dropout probability $ p = 0.5 $ is used in dense layers, mini-batch size of 50, and the initialization of parameters suggested by He et al. \cite{he2015delving}. A cross-entropy optimization with stochastic gradient descent is used, with a learning rate 0.04 that is reduced by half every 100,000 iterations. In this work, the Deep-Hits model was implemented in Tensorflow \cite{abadi2016tensorflow}. An initial process of 100,000 iterations of training is performed. After the first 100,000 iterations, an early-stopping criterion is tested on the validation set every 10,000 iterations. If the error on the validation set drops more than 1\% after 10,000 iterations, then the total number of iterations is extended in another 100,000 iterations from the current step. Otherwise, the training is stopped.

%If there is a sustained decrease in the validation error at the end of the initial 100,000 iterations, another training period of 100,000 iterations is added. \er{Furthermore, if at the end of the new 100,000 iterations the validation error does not decrease by at least a 99\% in relation to the previous period, the training is stopped assuming that the model has converged.}
\begin{table}[tb]
\centering
\small
\caption{Deep-HiTS architecture.}
\label{table:architecture}
\begin{tabular}{|c|c|c|}
\hline
\textbf{Layer}                                                   & \textbf{Layer Parameters} & \textbf{Output Size}                                    \\ \hline
Input                                                            & -                         & 21$\times$21$\times$4                                                 \\ \hline
Zero padding                                                     & -                         & 27$\times$27$\times$4                                                 \\ \hline
Rotation augmentation                                                   & -                         & 27$\times$27$\times$4                                                 \\ \hline
Conv.                                                            & 4$\times$4, 32                   & 24$\times$24$\times$32                                                \\ \hline
Conv                                                             & 3$\times$3, 32                   & 24$\times$24$\times$32                                                \\ \hline
Max-pool                                                         & 2$\times$2, stride 2             & 12$\times$12$\times$32                                                \\ \hline
Conv.                                                            & 3$\times$3, 64                   & 12$\times$12$\times$64                                                \\ \hline
Conv.                                                            & 3$\times$3, 64                   & 12$\times$12$\times$64                                                \\ \hline
Conv.                                                            & 3$\times$3, 64                   & 12$\times$12$\times$64                                                \\ \hline
Max-pool                                                         & 2$\times$2, stride 2             & 6$\times$6$\times$64                                                  \\ \hline
Flatten                                                          & -                         & 2304                                                    \\ \hline
\begin{tabular}[c]{@{}c@{}}Rotation\\ concatenation\end{tabular} & -                         & \begin{tabular}[c]{@{}c@{}}9216\\ (4$\times$2304)\end{tabular} \\ \hline
Dense                                                            & 9216$\times$64                   & 64                                                      \\ \hline
Dense                                                            & 64$\times$64                     & 64                                                      \\ \hline
Output softmax                                                   & 64$\times$2                      & 2                                                       \\ \hline
\end{tabular}
\end{table}

\subsection{Rotational Invariance}

Conventional CNN architectures include convolution layers to achieve translational invariance: a feature shifted to a different position at the input will have a similar representation at the output than the original feature. In addition, pooling operations aim at obtaining translational invariance at the local level. Many types of data exhibit rotational invariance properties, e.g. star and galaxy images. 

Some authors have attempted to directly encode rotational invariance in the architecture of a CNN \cite{cohen2016group}, \cite{dieleman2015rotation}, \cite{dieleman2016exploiting}. 
In particular Dieleman et al. \cite{dieleman2016exploiting}, exploit symmetries, by generating cyclic transformations of data, defined as counter clock-wise rotations of type $k\cdot90^{\circ},\,k\in\{0,1,2,3\}$. The first operation is to add a cyclic slicing layer at the input. This is done by stacking 4 cyclically rotated copies of each input sample in a batch, in such a way that if the original batch has size $N$, after the cyclic slicing layer the batch size will increase to $4N$. The first $N$ batch samples correspond to the original image, the following $N$ batch samples correspond to rotations in 90$^{\circ}$, and then 180$^{\circ}$ and 270$^{\circ}$. Mathematically, this operation is defined by $S(x)=[x, rx, r^2x, r^3x]^T$, where $r$ corresponds to the rotation operation of 90$^{\circ}$ in counter clock-wise direction, $x$ is the input batch to the layer. Column vectors are used to represent that the resulting feature maps are stacked in the batch dimension. With this operation, four images are generated, which are processed independently by the rest of model layers. The cyclic slicing layer is used by Dieleman et al. in \cite{dieleman2015rotation} to obtain rotation invariant CNNs for galaxy morphology prediction, and at the first layer of Deep-HiTS \cite{cabrera2017deep} to obtain partial rotational invariance for astronomical transients.

A second operation is the cyclic pooling layer, which combines the activations from the four rotated copies using a permutation-invariant function. The size of the mini-batch is reduced by 4. Formally, cyclic pooling is defined by an operation over the input $\textbf{x} = [x_0, x_1, x_2, x_3]^T$ as $P(\textbf{x}) = p (x_0, r^{-1}x_1, r^{-2}x_2, r^{-3}x_3)$, where $p$ corresponds to the pooling operation, e.g. average pooling is used in this work, which is applied to each unrotated minibatch. When applying cyclic pooling after a fully-connected layer, it becomes unnecessary to realign the features, since they lose their spatial structure. Although a cyclic pooling layer can be introduced in any part of the model after the cyclic slicing layer, in practice it is used before the output layer to obtain a rotation invariant network.

In \cite{dieleman2016exploiting} the authors introduced two additional operations: cyclic rolling and cyclic stacking. Both changes the number of feature maps, but the later also affects the batch size. In our experiments we did not find effective results using these operators, so they are not used in this work.

\subsection{Visualizing and Understanding CNN Decisions}

Recently several methods have been developed for understanding and visualizing what a CNN has learned in a classification task. 
In \cite{simonyan2013deep} a sensitivity analysis by computing gradients through backpropagation is proposed, yielding a saliency map. This method does not give a direct explanation of the network's score, but rather indicates which elements of an input would have to be modified and in which direction to make it belong more or less to the class decided by the CNN. In \cite{zeiler2014visualizing} a deconvolution strategy is proposed, which allows generating visualizations of a CNN at any intermediate layer. This method is exclusively limited to the convolutional layers of a network, where the main operations such as reverse filtering (un-filtering) and un-pooling, are similar to those used when propagating the gradient with back-propagation. 

A classification model can be seen as a function that maps a series of inputs $\textbf{x}$ to an output score $f(\textbf{x})$, which represents the confidence value of certain class membership. To interpret the decision of the network encoded in its parameters, operations that propagate $f(\textbf{x})$ backwards to the input space can be used. This approach generates a relevance value $R_i$ for each feature of $\textbf{x}$. Relevances quantify how much the respective features contribute to the value of the prediction score $f(\textbf{x})$. Positive values of $R_i$ are interpreted as features that contribute to the decision of the network, while negative values are interpreted as evidence against the prediction, potentially decreasing the value of $f(\textbf{x})$. Layer-wise relevance propagation (LRP) \cite{bach2015pixel} backpropagates the output of the network (not the gradients), layer by layer, towards the input space. As a result, we obtain the elements of the input that contribute in a positive or negative way to the classification score $f(\textbf{x})$, which will have respective positive or negative relevance values. LRP is based on the principle of the \emph{conservation of relevances}, i.e., that the sum of the relevances $R_j^{(l+1)}$ in a layer $l+1$ must be equal to the sum of the relevances $R_i^{(l)}$ in the previous layer $l$. Formally the conservation rule is expressed as follows:
\begin{equation}
\sum_{i}{R_i^{(l)}}=\sum_{j}{R_j^{(l+1)}}.
\label{eq:conservation_rule}
\end{equation}
Many mathematical operations can be created to backpropagate $f(\textbf{x})$ so that its intermediate layers satisfy Eq. (\ref{fig:CAP}), the simplest one is to propagate proportionally to the activations coming from previous neurons, in the same way as in a circuit node the electrical current is divided proportionally to the resistance of each branch\cite{montavon2017methods}, according to Ohm's and Kirchoff's laws. In this way we obtain the following equation:
\begin{equation}
R_{i\leftarrow j}^{(l,l+1)}=\frac{z_{ij}}{z_j}R_j^{(l+1)}.
\label{eq:prop_rule_1}
\end{equation}
The relevance $R_{i\leftarrow j}^{(l,l+1)}$ propagated from the neuron $j$ in the $l+1$ layer to the neuron $ i $ in the $ l $ layer, is equivalent to the ratio between the weighted activation $ z_{ij} = a_iw_{ij} $ of the neuron $ i $, and the sum of all forward pass activations coming from the $l$ layer $ z_j = \sum_{j}{z_{ij}} + b_j $ to the neuron $ j $, plus a bias. Although the addition of the bias breaks the conservation principle of Eq. (\ref{eq:conservation_rule}), it remains an acceptable approximation. Numerical instabilities are met when rule (\ref{eq:prop_rule_1}) is used, if $z_j\rightarrow 0$ then $R_{i\leftarrow j}\rightarrow \infty$, to avoid this situation a stabilizer hyperparameter  $\epsilon \geq 0$ is added, establishing the \emph{epsilon} rule as follows:
\begin{equation}
R_{i\leftarrow j}^{(l,l+1)}=\frac{z_{ij}}{z_j+\epsilon \, sign(z_j)}R_j^{(l+1)}.
\label{eq:prop_rule_e}
\end{equation}
Another propagation rule, that has no relevance loss by a stabilizer, is the $\alpha\beta$ rule, which separates positive activations, weights and biases $z_{ij}^+$ from negative ones $z_{ij}^-$, as follows:
\begin{equation}
R_{i\leftarrow j}^{(l,l+1)}=\left(\alpha\,\frac{z_{ij}^+}{z_j^+}-\beta\,\frac{z_{ij}^-}{z_j^-}\right)R_j^{(l+1)}.
\label{eq:prop_rule_ab}
\end{equation}
In Eq. (\ref{eq:prop_rule_ab}), it must be fulfilled that $ \alpha-\beta = 1 $ in such a way that the conservation rule is satisfied layer by layer. When naming this rule, the values used for parameters $\alpha$ and $\beta$ will define the model, for example, for  $\alpha=2$ and $\beta=1$, the method is called LRP-$\alpha_2\beta_1$. The current implementation of LRP considers layers with convolutional operations, max-pooling, avg-pooling, drop-out, multiple activations and fully-connected layers.
 \section{Enhanced Rotational Invariant CNN for Supernovae Detection}

The aim of this work is to enhance the CNN models obtained in our previous works for SNe detection \cite{cabrera2016supernovae}, \cite{cabrera2017deep}. For this purpose, Deep-Hits is taken as a baseline. There are three major contributions. The first one is to enhance the rotational invariant capability of Deep-HiTS by adding a cyclic pooling average layer. The second is to test ensemble classifiers. The third is to use the LRP method to analyze and visualize the CNN decisions, in particular to assess the rotational invariance property. The details are described in the Experiments section. The data used to train Deep-HiTS and the new model is explained in what follows.

\subsection{Data}

Since SNe are extremely rare events, negative samples (artifacts) are obtained by running the HiTS pipeline on the 2013 survey, where 40 sky fields were observed in the $u$ band, approximately every 2 hours for 4 consecutive nights. The negative samples are caused mostly by background fluctuations, interference with other objects, poor astrometry, and defective CCD pixels. A total of 802,087 negative samples were generated\footnote{Data-set may contain some transients not present in the reference image, but we conservatively estimate them as a 0.2\% of total data.}. As SNe are rare events, in order to get a balanced dataset, positive candidates were simulated by picking stars from actual observations, applying a specific SNR distribution and then inserting them back into the \emph{science} image in a different location with respect to the original source. As explained above, each sample consists of four 21$\times$21 images, that are stacked. Out of 1,604,174 data samples available, a fixed amount of 1,220,000 was selected for training, 100,000 for validation and 100,000 for testing. This dataset, as well as the Deep-HiTS code, is publicly available at the following link https://github.com/guille-c/Deep-HiTS. We use accuracy, precision, recall, and f1 score as performance measures, and we plot detection error trade-off curves (DET), depicting FNR versus FPR, to evaluate the quality of models at different operation points.

\section{Experiments}

\subsection{Enhancing Deep-HiTS}

\begin{figure*}[h!]
\centerline{\includegraphics[width=1\textwidth]{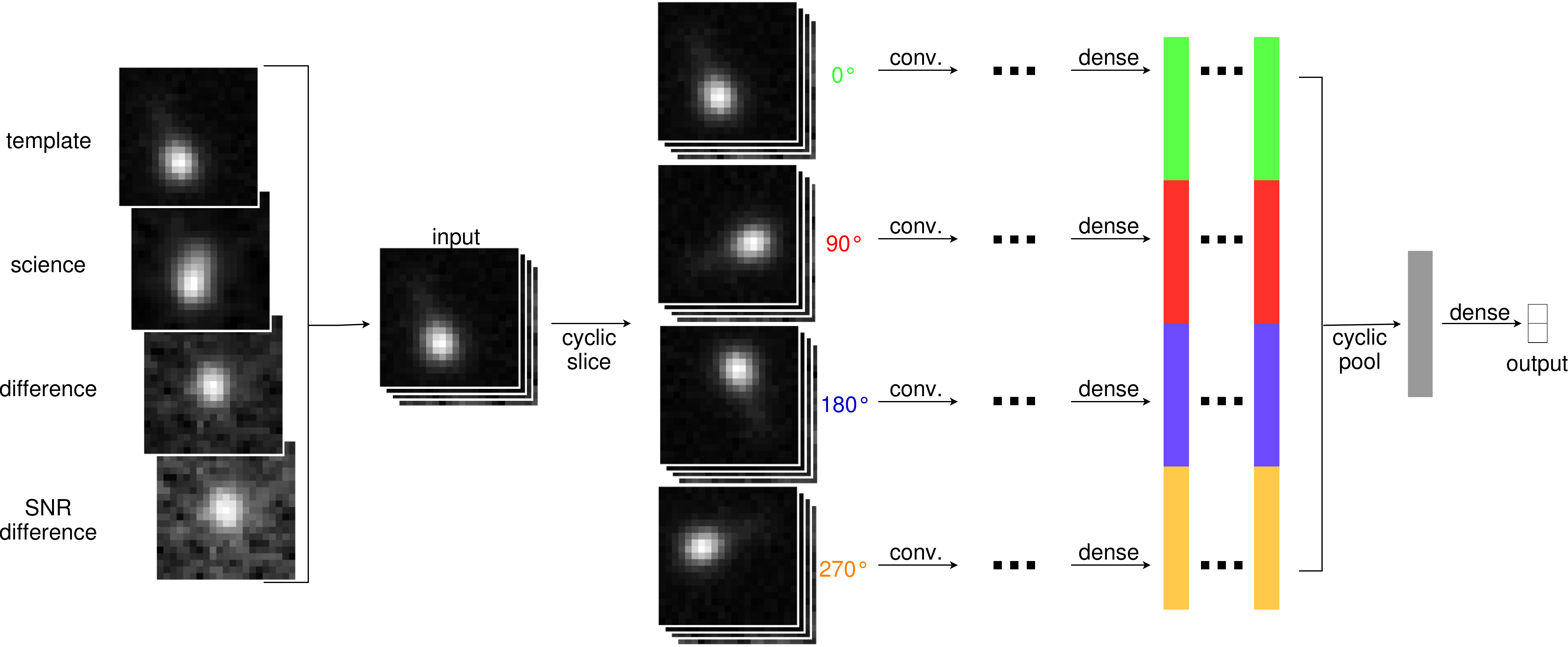}}
\caption{Schematic architecture of CAP that represents the effect of cyclic slice and cyclic pooling layers. It shows a four channels input that passes through the cyclic slicing layer where rotation copies of it are stacked along the batch dimension. Afterwards, each rotation is processed by convolutional and dense layers independently, until reaching the cyclic pooling layer, where all rotation features (color coded) are fused together to be processed by the output layer.}
\label{fig:CAP}
\end{figure*}

We tested cyclic symmetric operations in order to improve the rotational invariance property of Deep-HiTS. The cyclic slicing is maintained since it is exactly the same operation of rotation augmentation used by the original Deep-HiTS model. In a variant to Deep-HiTS, we implemented the cyclic average pooling (CAP) operation just before the output layer. This means that instead of reordering features prior to the dense layers, as Deep-HiTS does, the features associated with each rotation continue to be processed independently of each other. The CAP operation is applied before the output layer of the network, so that, for each sample, the features coming from each rotation are averaged, thus the output score of the network will always be the same, regardless of whether there is a cyclic rotation in the input. This model is proposed based on the prior knowledge that the detection of SNe is independent of angular rotations of samples. The proposed model with cyclic average pooling will be referred as CAP from now on, and it can be seen in Fig. \ref{fig:CAP}. The CAP architecture is identical to Deep-HiTS until the last convolutional layer. We also changed to ReLU activations instead of leaky ReLUs, for simplicity in the models. The first two rows of Table \ref{table:measures} show that there is no significant difference between the model with leaky ReLUs versus ReLUs. Therefore ReLU activations were used for the rest of the architectures tried in this paper.

% TABLE: PERFORMANCEE METRICS
\begin{table*}[t]
\centering
\normalsize
\caption{Models performance over test set. CAPE shows better results in all metrics with respect to DH. Welch's t-test p-values over DH and CAP results show statistical significance, having values below 10$^{-2}$ (except for recall). Applying the same test over CAP and CAPE show no statistical significance, which means there is no major difference using one or another.}
\label{table:measures}
\begin{tabular}{ccccc}
\hline
Model                                                                                                        & Accuracy       & Precision      & Recall         & F1-score       \\ \hline
\begin{tabular}[c]{@{}c@{}}Deep-Hits (DH)\end{tabular}                                               & 99.45$\pm$0.02 & 99.37$\pm$0.04 & 99.55$\pm$0.06 & 99.45$\pm$0.02 \\ 
Deep-Hits + ReLU                                                                                             & 99.47$\pm$0.02 & 99.44$\pm$0.04 & 99.52$\pm$0.06 & 99.47$\pm$0.02 \\ \hline
\begin{tabular}[c]{@{}c@{}}Cyclic Avg. Pool (CAP)\end{tabular}                                     & 99.52$\pm$0.01 & 99.45$\pm$0.01 & 99.61$\pm$0.02 & 99.52$\pm$0.01 \\ 
\textbf{\begin{tabular}[c]{@{}c@{}}Cyclic Avg. Pool Ensemble (CAPE)\end{tabular}} & \textbf{99.53$\pm$0.01} & \textbf{99.45$\pm$0.02} & \textbf{99.63$\pm$0.03} & \textbf{99.53$\pm$0.01} \\ \hline \hline 
\begin{tabular}[c]{@{}c@{}}Welch's t-test p-value\\ DH v/s CAP\end{tabular}                   & 2$\times$10$^{-5}$  & 4.2$\times$10$^{-3}$  & 6.6$\times$10$^{-2}$  & 2.2$\times$10$^{-5}$  \\ \hline
\begin{tabular}[c]{@{}c@{}}Welch's t-test p-value\\ CAP v/s CAPE\end{tabular}                   & 1.8$\times$10$^{-1}$  & 8.9$\times$10$^{-1}$  & 2.4$\times$10$^{-1}$  & 1.8$\times$10$^{-1}$  \\ \hline
\end{tabular}
\end{table*}

%FIGURE: ARCHITECTURE

Table \ref{table:measures} shows the means and standard deviations obtained in 6 trials with different initializations for the following measures: accuracy, precision, recall, and f1-score for each model. An improvement of all metrics is observed when introducing the layer of cyclic pooling average, with respect to Deep-HiTS. According to the Welch's hypothesis test, all CAP measures (except recall) show statistical significance with a probability of less than 10$^{-2}$ that these results were obtained by chance. A CAP ensemble (CAPE) of 3 individual models was generated, with the objective of obtaining better performances. The majority rule was applied to combine the classifiers. Table \ref{table:measures} shows that CAPE obtains results that are indistinguishable from CAP, i.e. there are no statistically significant differences. 

\begin{figure*}[t]
\centering
\begin{subfigure}{.48\textwidth}
  \centering
  \includegraphics[width=1\linewidth]{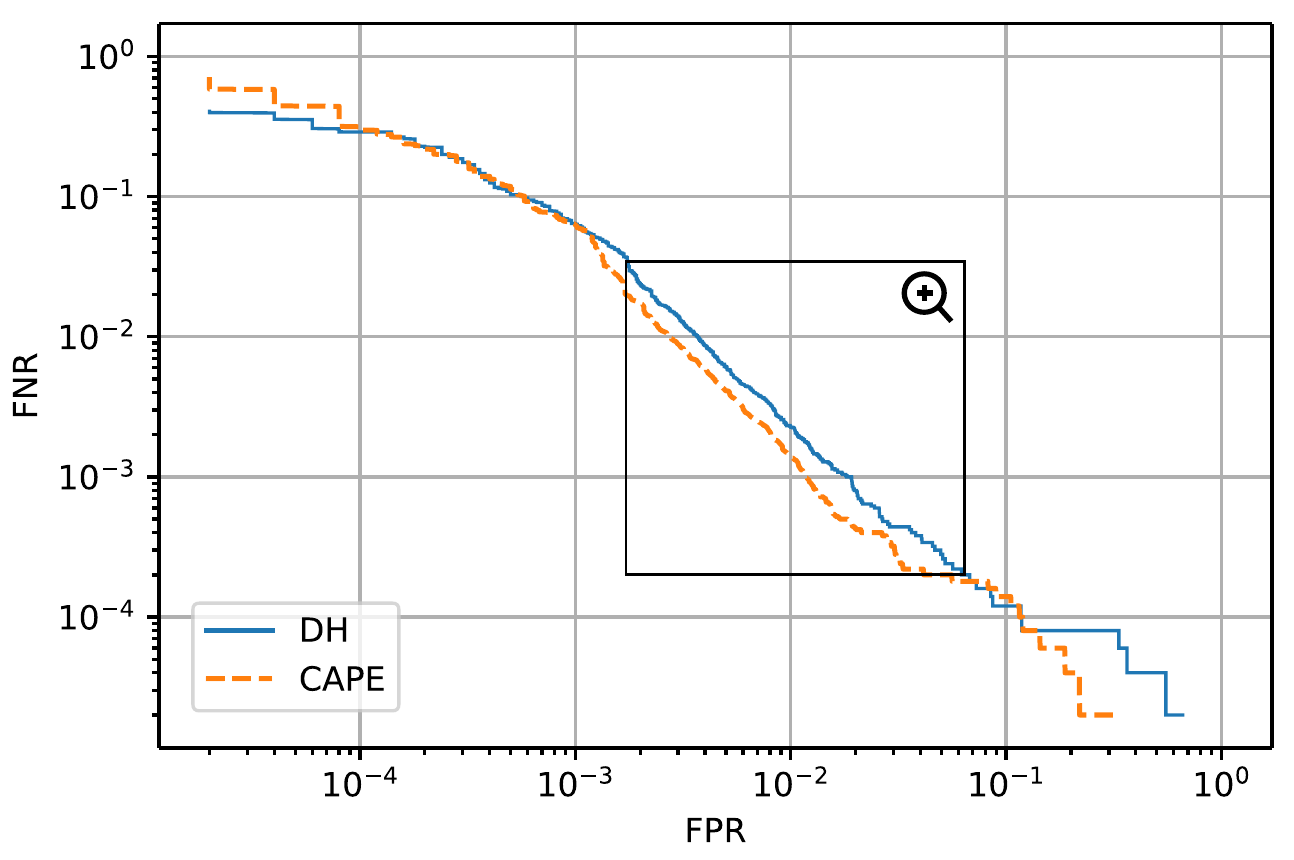}
  \caption{}
  \label{fig:DETnormal}
\end{subfigure}%
\begin{subfigure}{.5\textwidth}
  \centering
  \includegraphics[angle=-90, width=1\linewidth]{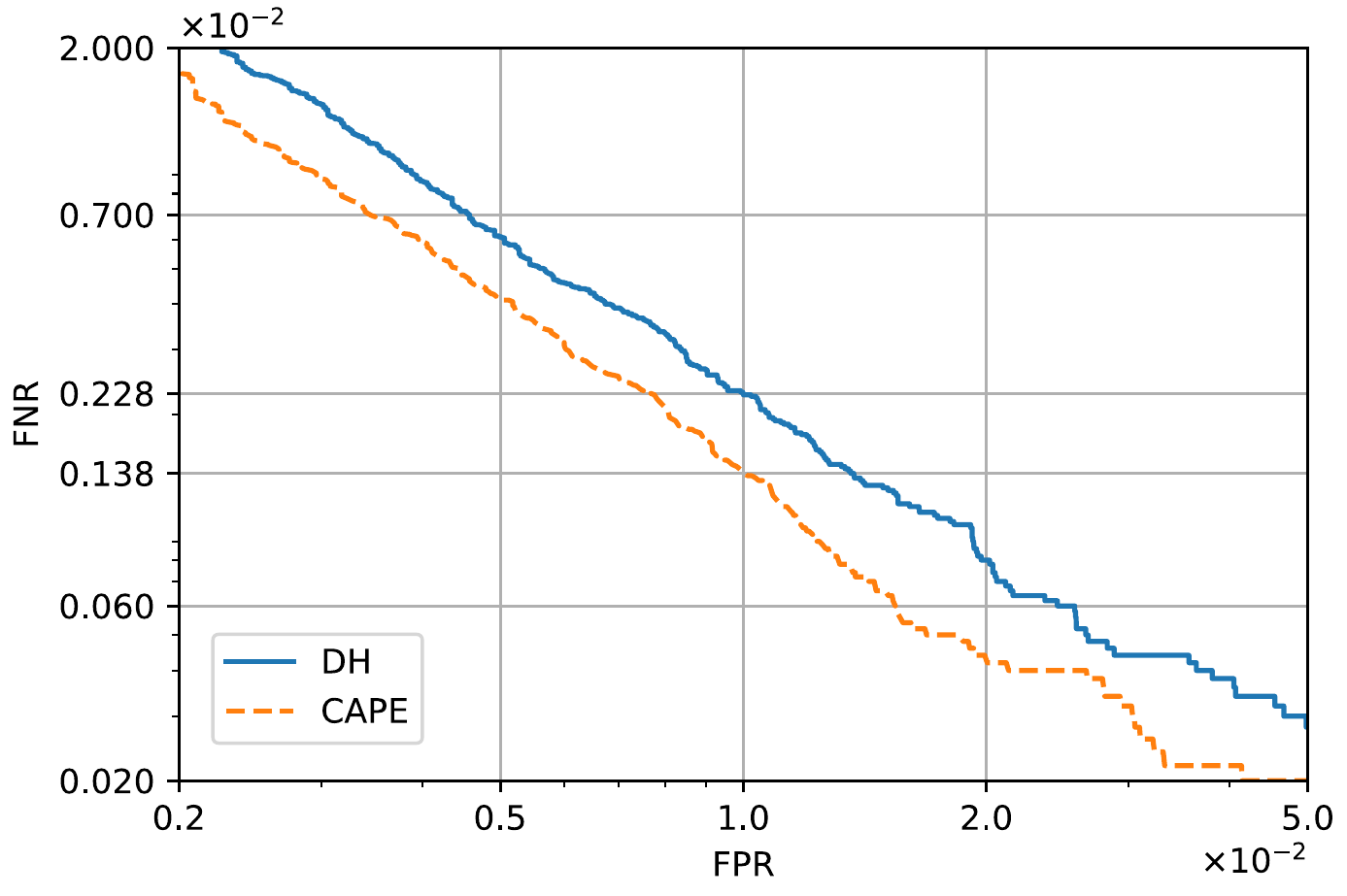}
  \caption{}
  \label{fig:DETzoom}
\end{subfigure}
\caption{Detection error tradeoff (DET) curves of DH and CAPE. While a curve is closer to the bottom left corner, the better the model it represents. (a) Whole range DET curves, with a black box to be zoomed in (b). (b) Zoom of (a) DET curves, which shows a $\sim$40\% improvement in FNR of CAPE with respect to DH, around FPR 10$^{-2}$.}
\label{fig:DET}
\end{figure*}

The performances of Deep-HiTS and CAPE can be compared throughout different operation points, by calculating their detection error tradeoff (DET) curves, which plots false negative rate (FNR) versus false positive rate (FPR), see Fig. \ref{fig:DET}. Better models present lower FNR and FPR with a DET curve near the lower left corner. The DET curve corresponding to CAPE is below the Deep-HiTS curve for a large range of user detection thresholds. To better appreciate the difference, a zoom around the operation point FPR $\sim$10$^{-2}$ is shown in Fig. \ref{fig:DETzoom}. This operation point is used in Deep-Hits to make comparisons. When FPR $\sim$10$^{-2}$, the proposed model reaches an FNR of 1.38$\times$10$^{-3}$, while Deep-HiTS gets an FNR of 2.28$\times$10$^{-3}$.

% FIGURE: DET CURVES

% FIGURE: SNe CANDIDATE DETECTED AS BOGUS BY DH
\begin{figure*}[h!]
\centering
\begin{subfigure}{.5\textwidth}
  \centering
  \includegraphics[width=0.78\linewidth]{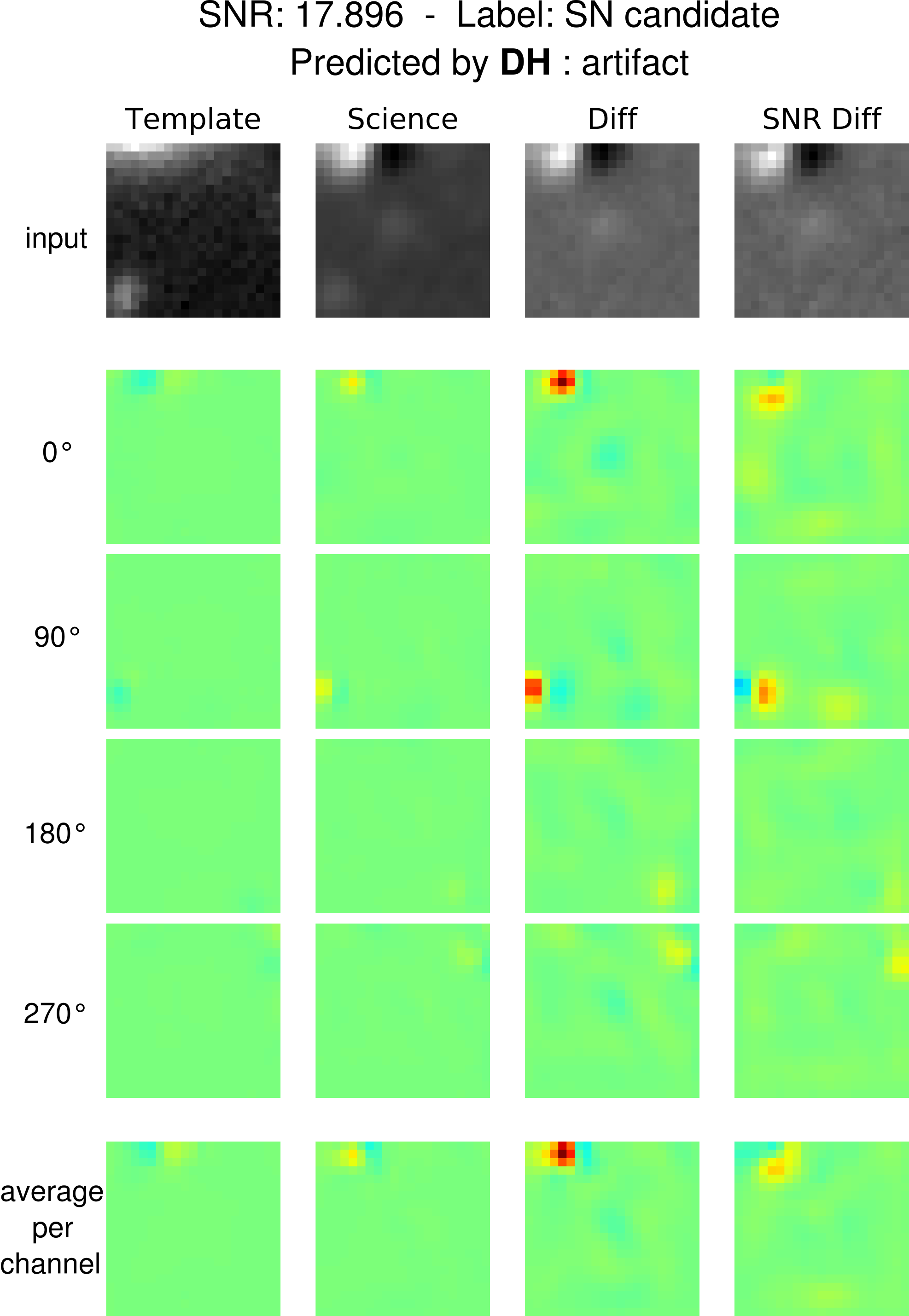}
  \caption{}
  \label{fig:DHbogus}
\end{subfigure}%
\begin{subfigure}{.5\textwidth}
  \centering
  \includegraphics[width=0.78\linewidth]{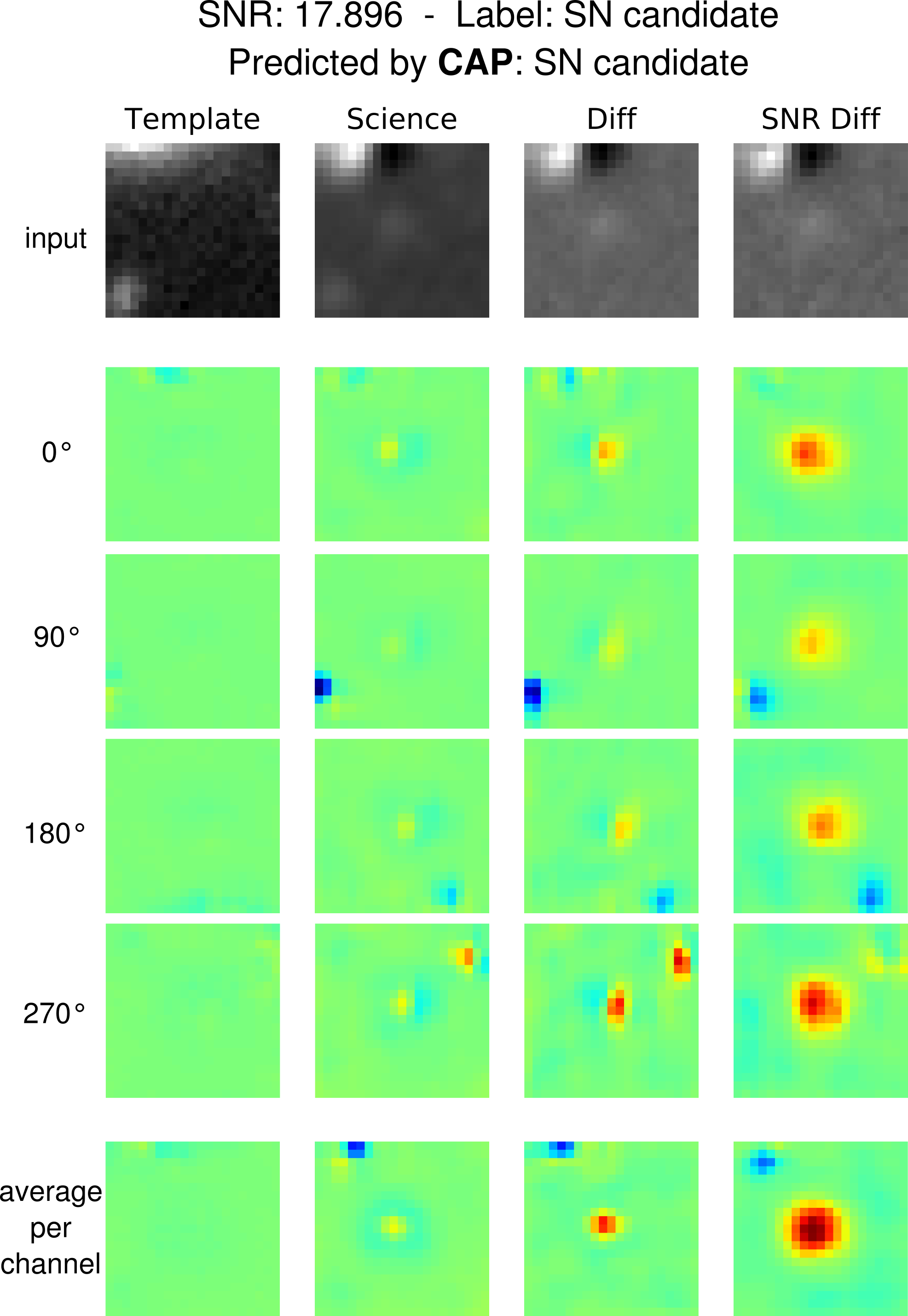}
  \caption{}
  \label{fig:CAPreal}
\end{subfigure}
\caption{LRP-$\alpha_2\beta_1$ heatmaps for DH and CAP, when propagating output score of predicted class for each model. Sample used is an `SN candidate' that is misclassified by DH and correctly classified by CAP. LRP relevance heatmaps are shown for each rotation input, and then their average per channel when rotations are realigned. (a) DH heatmaps. (b) CAP heatmaps.}
\label{fig:LRP}
\end{figure*}

% FIGURE: BOGUS DETECTED AS REAL IN DH
\begin{figure*}[h!]
\centering
\begin{subfigure}{.5\textwidth}
  \centering
  \includegraphics[width=0.78\linewidth]{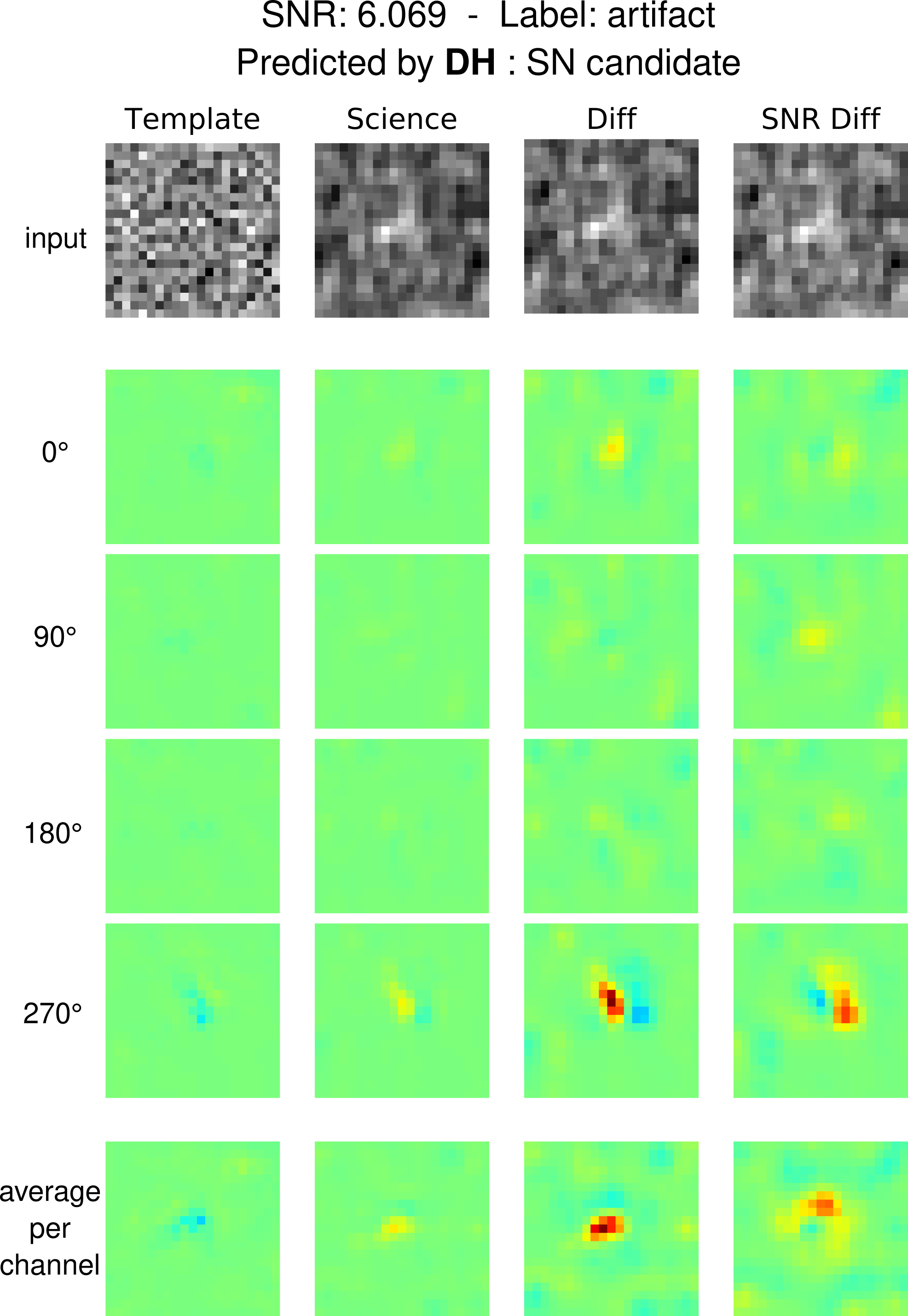}
  \caption{}
  \label{fig:DHreal}
\end{subfigure}%
\begin{subfigure}{.5\textwidth}
  \centering
  \includegraphics[width=0.78\linewidth]{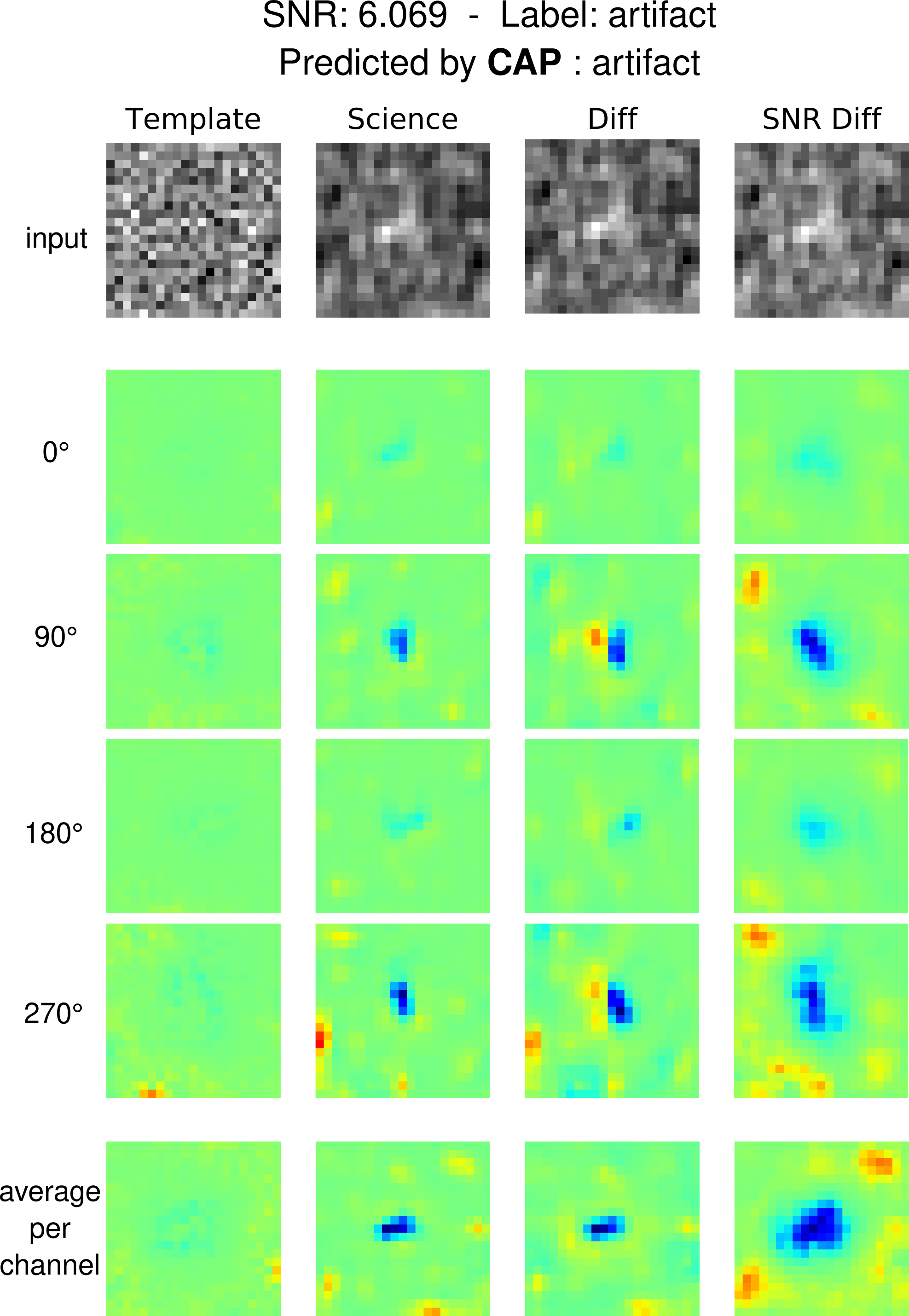}
  \caption{}
  \label{fig:CAPbog}
\end{subfigure}
\caption{LRP-$\alpha_2\beta_1$ heatmaps for DH and CAP, when propagating output score of predicted class for each model. Sample used is an `artifact' that is misclassified by DH and correctly classified by CAP. LRP relevance heatmaps are shown for each rotation input, and then their average per channel when rotations are realigned. (a) DH heatmaps. (b) CAP heatmaps.}
\label{fig:LRPimpBog}
\end{figure*}

% FIGURE: CORRECT DETECTION OF A SNe BY DH AND CAP
% \begin{figure*}[h!]
% \centering
% \begin{subfigure}{.5\textwidth}
%   \centering
%   \includegraphics[width=0.78\linewidth]{RealCorrectDH}
%   \caption{}
%   \label{fig:sub1}
% \end{subfigure}%
% \begin{subfigure}{.5\textwidth}
%   \centering
%   \includegraphics[width=0.78\linewidth]{RealCorrectPM}
%   \caption{}
%   \label{fig:sub2}
% \end{subfigure}
% \caption{LRP-$\alpha_2\beta_1$ heatmaps for DH and CAP, when propagating output score of predicted class for each model. Sample used is a `artifact' correctly classified by each model. Heatmaps relevances are shown for each rotation input, and then their per channel mean when rotations are realigned. (a) DH heatmaps. (b) CAP heatmaps.}
% \label{fig:LRPrealCorrectDH}
% \end{figure*}

\subsection{LRP Analysis and Visualization}

We use the LRP method to visualize the effect of rotational invariance provided by the CAP model, and its advantages over the Deep-HiTS model. The rotational invariance property is reflected when projecting the relevances to the rotations within each channel. The LRP method needed to be adapted to each model. For Deep-HiTS, the reverse operation of the forward step for the feature reordering layer is simply carried out, to change from the reordered features to the feature maps propagated by the convolutional layers. In the case of CAP, the current implementation of LRP is adapted to reorder the features to perform a 2$\times$2 average-pooling, where each element of the filter is a feature associated with a different rotation. In this way, the propagation step of relevances through the CAP layer becomes the same as a normal average-pooling layer.

First, we used the LRP visualization method to analyze 197 test samples where CAP gave correct predictions, while Deep-HiTS made wrong decisions. Fig. \ref{fig:LRP} shows the case of a sample labeled `SN candidate', and Fig. \ref{fig:LRPimpBog} shows the case of a sample labeled `artifact'. Both Figs. show on the top row the 4 source images (channels). The next four rows display the heatmaps corresponding to cyclic rotations of $k\cdot90^{\circ},\, k\in\{0,1,2,3\}$. The bottom row depicts the average of the unrotated heatmaps per channel. These heatmaps represent LRP relevances when propagating network's prediction scores with the $\alpha_2\beta_1$ rule. For display purposes the heatmaps were normalized between 0-1 for each sample, so that they are comparable to each other. 

The heatmaps of Fig. \ref{fig:DHbogus} show that the relevance is concentrated in the upper left corner of the unrotated \emph{difference} image. Our interpretation is that the Deep-HiTS decision of class `artifact' is supported by the light source observed in that region of the \emph{difference} image. In contrast, the heatmaps of Fig. \ref{fig:CAPreal} show that the relevance is concentrated in the center of the \emph{difference} and \emph{SNR difference} images, providing evidence of the presence of an SN. The light source that confuses the Deep-HiTS model appears in blue in the CAP heatmaps, i.e., as negative relevances indicating that the presence of such a light source decreases the prediction score of the SN. 

A property of the CAP model is that the heatmaps corresponding to cyclic rotations are more evenly distributed. Compare for example the last columns (SNR diff) in Figs. \ref{fig:DHbogus} and \ref{fig:CAPreal}, or the last columns in Figs. \ref{fig:DHreal} and \ref{fig:CAPbog}. More uniform distributions of heatmaps for cyclic rotations is an indication that the model is more rotational invariant.

Likewise Fig. \ref{fig:LRPimpBog} corresponds to an `artifact' sample. The blue color in \emph{science}, \emph{difference} and \emph{SNR difference} in CAP heatmaps is more evenly distributed across rotations, indicating that the bright source in the middle of the images reduces the prediction score because it looks similar to an SN candidate. On the other hand, Deep-HiTS classifies the same sample as `SN candidate', based on the evidence presented at the center of the 270$^{\circ}$ rotated \emph{difference} and \emph{SNR difference} images. 

% In order to analyze correctly classified samples by DH, Fig. \ref{fig:LRPrealCorrectDH} shows same heatmaps as before, but for a `SN candidate' sample. In Fig. \ref{fig:LRPrealCorrectDH} can be seen how highest relevances for DH prediction are found in the 270$^{\circ}$ \emph{difference} and \emph{SNR difference} images, while for CAP they appear more evenly through all rotations.

The qualitative interpretation of the heatmaps presents some ambiguity. The red color stands for positive relevances in favor of the CNN's prediction, and the blue color stands for negative evidence against the network's decision. However, each heatmap color can be interpreted as the presence or absence of the respective input feature. For example, the blue color located over the center in Fig. \ref{fig:LRPimpBog}b is an indicator that the brightness diminishes the output confidence. But blue color corresponding to image regions with darker pixels indicates that the model is getting lower prediction scores due to an absence of brightness.

To validate quantitatively the conjecture that the CAP model is more rotational invariant than Deep-HiTS, we computed a measure called \emph{Average Standard Deviation of Cyclic Rotations} ($\bar{\sigma}$) using 5,000 test samples. This measure is computed per channel, so there are four different values per sample, corresponding to each of the 4 channels. To compute $\bar{\sigma}$, we need first to choose a channel $c$ and its $i-th$ rotation $c_i$, then calculate the pixel-wise variance for its LRP-$\alpha_2\beta_1$ heatmap $h^{c_i}$. This is done by computing:
\begin{equation}
Var(h^{c_i}) = \frac{\sum_{j=1}^{N}(h_j^{c_{i}}-\mu_{j}^c)^2}{N},
\label{eq:hm_var}
\end{equation}
where $h_j^{c_{i}}$ is the $j-th$ pixel, and N is the total number of pixels in a heatmap, in this case $N=441$. The variance is computed with respect to the average heatmap per channel $\mu_{j}^c$. Examples of average heatmaps can be found at the bottom row in Figs. \ref{fig:LRP} and \ref{fig:LRPimpBog}. Finally, $\bar{\sigma}$  is calculated by computing the root square of $Var(h^{c_i})$ for each rotation heatmap, and then averaging over the four rotations, as follows:
\begin{equation}
\bar{\sigma} = \frac{\sum_{i=1}^{N_r}\sqrt{Var(h^{c_i})}}{N_r},
\label{eq:hm_sigma}
\end{equation}
where $N_r=4$ is the number of rotations at the input of the CNN.

\begin{figure}
\centering

\begin{subfigure}[b]{0.48\textwidth}
   \includegraphics[width=1\linewidth]{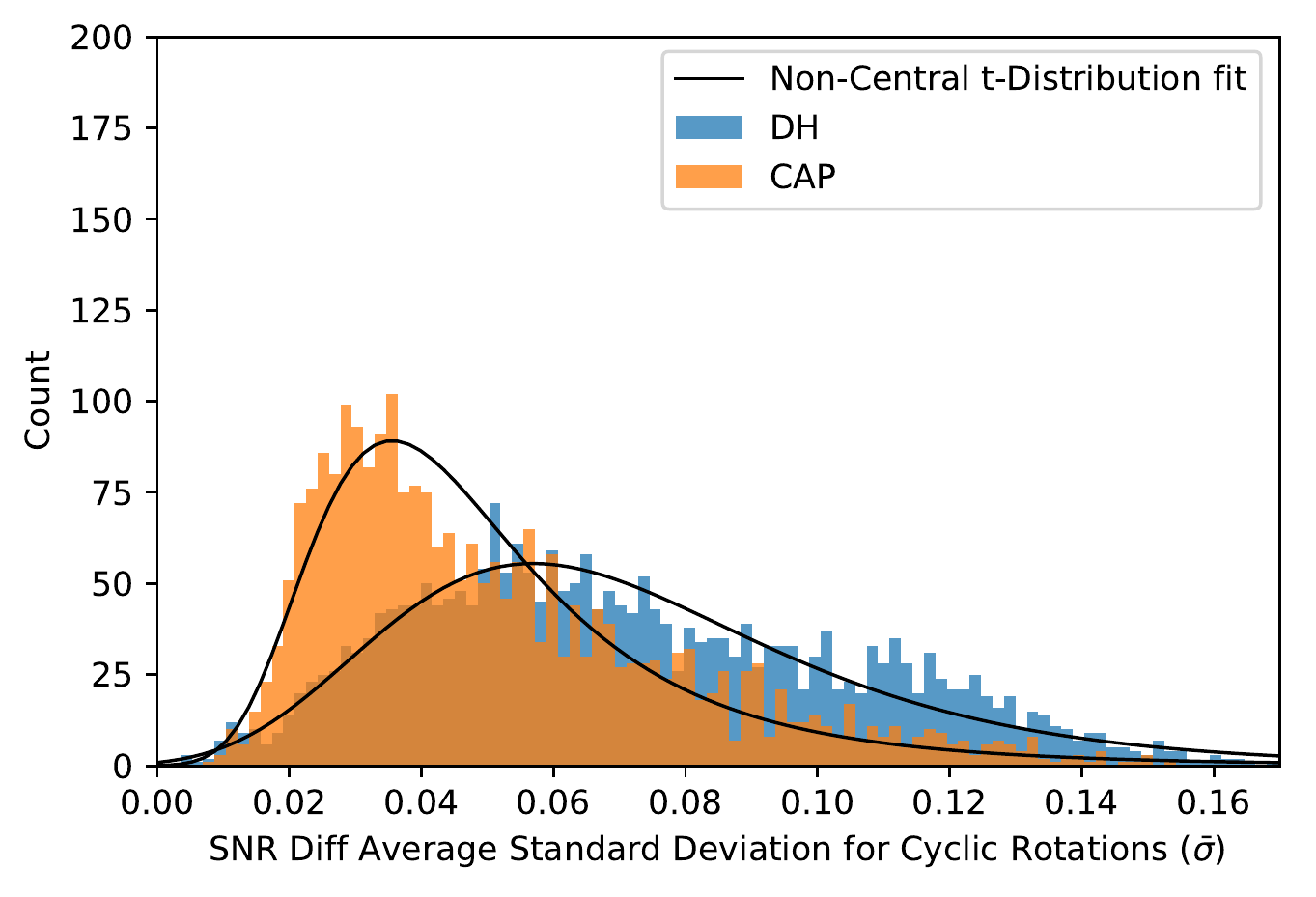}
   \caption{}
   \label{fig:histArt}
\end{subfigure}

\begin{subfigure}[b]{0.48\textwidth}
   \includegraphics[width=1\linewidth]{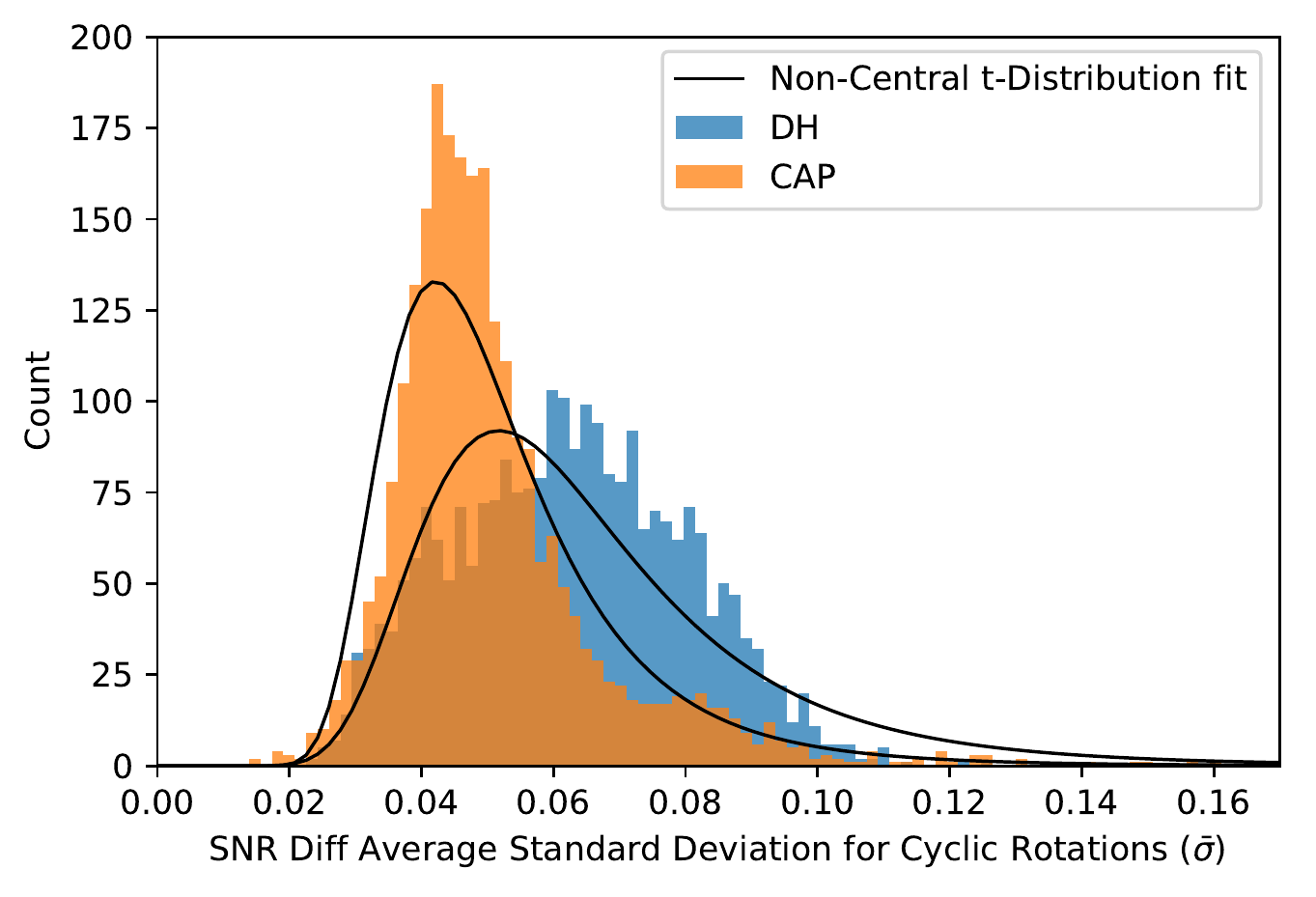}
   \caption{}
   \label{fig:histSN}
\end{subfigure}

\caption{Average Standard Deviation for Cyclic Rotations ($\bar{\sigma}$) histograms for `artifact' and `SN candidate' classes for the DH and CAP models. A \emph{Non-Central t-Distribution} is fitted to each histogram. (a) ($\bar{\sigma}$) histograms for `artifact' class. (b) ($\bar{\sigma}$) histograms for `SN candidate' class.}
\label{fig:hists}
\end{figure}

As the SNR difference image is the channel where transients are most significantly observed, we plot histograms of $\bar{\sigma}$ corresponding to \emph{SNR difference}, for both Deep-HiTS and CAP models. Fig. \ref{fig:hists} shows the histograms separated by class: artifacts and SN candidates. For the class `artifact' there are 2,499 samples and for the class `SN candidate' 2,501. To statistically compare the histograms generated for Deep-HiTS and CAP, a \emph{Non-Central t-Distribution} probability distribution is fitted to each histogram using the open source scientific tools library for Python (\emph{SciPy}, \cite{jones2014scipy}). The curve computed with \emph{SciPy} is normalized so that its integral is 1, thus the curve must be re-scaled to fit the original histogram. The scale factor is obtained by normalizing the histogram in the same way as the fitted distribution does. The normalized curve is multiplied by the scale factor in order to plot it over the histogram.

% FIGURE: FITED DISTRIBUTIONS OF STDs

Table \ref{table:plot Metrics} shows the mean, variance, skewness and kurtosis obtained from the fitted distributions to `artifact' and `SN candidate' classes. It can be observed that for both classes the CAP model has a lower mean and a larger kurtosis and skewness than the Deep-HiTS model. This effect is clearly observed in Fig. \ref{fig:histSN}, where the CAP histogram shows a leptokurtic distribution, as well as a lower mean value of $\bar{\sigma}$ with respect to the one of Deep-HiTS. This means that CAP relevances are distributed more evenly throughout cyclic rotations, while Deep-HiTS heatmaps tend to focus its relevances on specific rotations. This fact is supporting evidence that the CAP model is more rotational invariant than the Deep-HiTS model.

% TABLE: PER SAMPLE STD
\begin{table}[h]
\centering
\normalsize
\caption{Moments of Non-Central t-Distribution fitted to $\bar{\sigma}$ histograms of `artifact' and `SN candidate' classes for DH and CAP.}
\label{table:plot Metrics}
\begin{tabular}{cc|c|c|c|c|}
\cline{3-6}
                                                                                               &     & \begin{tabular}[c]{@{}c@{}}$\bar{\sigma}$\\ mean\end{tabular} & \begin{tabular}[c]{@{}c@{}}$\bar{\sigma}$\\ var\end{tabular} & \begin{tabular}[c]{@{}c@{}}$\bar{\sigma}$\\ skew\end{tabular} & \begin{tabular}[c]{@{}c@{}}$\bar{\sigma}$\\ kurt\end{tabular} \\ \hline
\multicolumn{1}{|c|}{\multirow{2}{*}{Artifacts}}                                               & DH  & 0.0729                                                & 0.0014                                                    & 0.38                                                              & -1.34                                                              \\ \cline{2-6} 
\multicolumn{1}{|c|}{}                                                                         & CAP & 0.0517                                                & 0.001                                                    & 1.08                                                              & -0.28                                                             \\ \hline
\multicolumn{1}{|c|}{\multirow{2}{*}{\begin{tabular}[c]{@{}c@{}}SN\\ candidates\end{tabular}}} & DH  & 0.0655                                                & 0.0007                                                       & 1.09                                                              & -0.28                                                              \\ \cline{2-6} 
\multicolumn{1}{|c|}{}                                                                         & CAP & 0.0519                                                & 0.0003                                                    & 1.62                                                               & 1.23                                                             \\ \hline
\end{tabular}
\end{table}

%%%%%%%%%%%%%%%%%%%%%%%%%%%%%%%%%%%%%%%%%%%%%%%%%%%%%%%%%%%%

\section{Conclusions and Future Work}

In this work, we enhanced the rotational invariant capability of the Deep-HiTS model by adding a cyclic pooling average layer. The results are consistent with the hypothesis that astronomical objects do not depend on the angle on which the image is observed given the same conditions of observation. An ensemble of CAP models obtained the best results so far with the HiTS dataset, reaching an average accuracy of 99.53\%. The improvement over Deep-HiTS is significant both statistically and in practice. For example, for a standard operation point with FPR $\sim$10$^{-2}$, the proposed model achieves an FNR of 1.38x10$^{-3}$, which entails a $\sim$40\% reduction of missing transients with respect to Deep-HiTS. From the astronomer viewpoint, it is important not to miss positive samples of rare SNe events.

We have used the LRP method to visualize and analyze the heatmaps showing the most relevant pixels for the discrimination task at hand. We defined a measure to assess quantitatively the rotational invariance capability of the different models. The results show that the proposed model is more rotational invariant than the original Deep-HiTS model. This is a novel application for the LRP method and the first time that it has been applied to astronomical data. 

LRP is a positive step towards understanding and visualizing what a CNN has learned. However, the tool may be improved to visualize intermediate layers, adding gradient information, as well as improving the interpretation of positive and negative relevances. Obtaining the best ensemble classifier can also be investigated by changing the size and the rule of the ensemble.

%In Fig. \ref{fig:LRP} it can be seen that red color comes mostly from the light source of the \emph{SNR difference} image's center since these are the main sources of network prediction score. However cases of correct class `artifact' classifications were found where positive relevances were also found in the center of \emph{difference} images, when there was no presence of a light source, the interpretation of this situation being the fact that the absence of light sources in the central zone contributes in greater extent to the confidence of the prediction `artifact'.

%%%%%%%%%%%%%%%%%%%%%%%%%%%%%%%%%%%%%%%%%%%%%%%%%%%%%%%%%%%%%%%%%%%%%%%%%%%%%%

\section{Acknowledgments}
Pablo Estévez, Pablo Huijse and Guillermo Cabrera-Vives acknowledge support from FONDECYT through grants 1171678, 1170305 and 3160747, respectively. Francisco F\"orster acknowledges support from CMM Basal Project PFB-03. The authors thanks the support of Conicyt through project DPI20140090. Ignacio Reyes acknowledges financial support from CONICYT-PCHA through its M.Sc. scholarship 2016 number 22162464. The authors acknowledge support from the Chilean Ministry of Economy, Development, and Tourism's Millennium Science Initiative through grant IC12009, awarded to the Millennium Institute of Astrophysics, MAS.

%%%%%%%%%%%%%%%%%%%%%%%%%%%%%%%%%%%%%%%%%%%%%%%%%%%%%%%%%%%%%%%%%%%%%%%%%%%%%%%%

\bibliographystyle{ieeetr}
\bibliography{bibliografia}
\end{document}